\documentclass[11pt]{article}
\newcommand{\be}{\begin{equation}}
\newcommand{\ee}{\end{equation}}
\newcommand{\bea}{\begin{eqnarray}}
\newcommand{\eea}{\end{eqnarray}}
\newcommand{\sn}{{\rm sn}}

\newcommand{\dn}{{\rm dn}}
\newcommand{\cn}{{\rm cn}}
\newcommand{\sech}{{\rm sech}}

\begin{document}

\vspace{0.5in}
\begin{center}
{\LARGE{\bf Some Novel Aspects of the Plane Pendulum in Classical Mechanics}}
\end{center}
\vspace{0.5in}
\begin{center}
{\LARGE{\bf Avinash Khare}} \\
{Physics Department, Savitribai Phule Pune University \\
 Pune 411007, India}
\end{center}

\begin{center}
{\LARGE{\bf Avadh Saxena}} \\ 
{Theoretical Division and Center for Nonlinear Studies, 
Los Alamos National Laboratory, Los Alamos, New Mexico 87545, USA}
\end{center}

\vspace{0.9in}
\noindent{\bf {Abstract:}}

We obtain a novel connection between the exact solutions of the plane pendulum, 
 hyperbolic plane pendulum and inverted plane pendulum equations as well as the 
 static solutions of the sine-Gordon and the sine hyperbolic-Gordon equations 
 and obtain a few exact solutions of the above mentioned equations. Besides, 
 we consider the plane pendulum equation in the first anharmonic approximation 
 and obtain its large number of exact periodic as well as hyperbolic solutions.
 In addition, we obtain two exact solutions of the plane pendulum equation in 
the second anharmonic approximation. Further, we introduce an elliptic plane 
pendulum equation in terms of the Jacobi elliptic functions 
-$\sn(\theta,m)/\dn(\theta,m)$ which smoothly goes over to the the plane 
pendulum equation in the $m=0$ limit and the hyperbolic plane pendulum equation
in the $m = 1$ limit where $m$ is the modulus of the Jacobi elliptic functions.
We show that in the harmonic approximation, the elliptic pendulum problem 
represents a one-parameter family of isochronous system. Further, for the 
special case of $m = 1/2$, we show that one has an isochronous 
system even in the first anharmonic approximation. Finally, we also briefly 
discuss the hyperbolic plane pendulum and obtain a few of its exact solutions 
in the harmonic as well as the first anharmonic approximation.

\section{Introduction}

The plane pendulum system is one of the celebrated problems in classical
mechanics and has been discussed in almost every classical mechanics book. 
While most of the books are content in terms of discussing its solution in the 
harmonic approximation, there are books which have also discussed some of its 
exact solutions \cite{cs}. Apart from it being an educational tool, the plane 
pendulum is useful in understanding oscillations and nonlinear oscillators. It 
serves as a model system for exploring topics such as chaos theory and the 
behavior of driven or damped pendulums. Many aspects of the 
plane pendulum problem in quantum mechanics are well understood \cite{con,pk} 
basically since it is a linear problem. For some recent papers on this subject
see \cite{dou,sch,lei,ayu,say,don,bak}. On the other hand, plane pendulum 
problem in classical mechanics continues to give new surprises. It 
involves exact solutions of a nonlinear problem and one is never sure if one 
has obtained all possible exact solutions of the plane pendulum equation. 
Most of the activity is concentrated in finding an accurate expression for the
time period of the plane pendulum away from the harmonic approximation 
\cite{gan,cod,mol,kid,par,hit,bel,lim,amo,xin,lim1,joh1,joh2}. It is
worth reminding that in principle the exact solution of the plane pendulum 
(and hence the time period) is known in terms of the Jacobi elliptic function 
\cite{con,pk}.

The purpose of the present paper is to shed some light on what we believe
to be a few new features of the plane pendulum problem. To begin with,
we show that with  suitable Ansatz the static solutions of the sine-Gordon 
(SG) and the sine hyperbolic-Gordon (SHG) equations as well as the 
(time-dependent) solutions of the inverted plane pendulum and the plane 
pendulum equations are all related to the solutions of a nonlinear equation. 
We then present five solutions of this nonlinear equation and hence  obtain 
five solutions of the above mentioned four equations.  Further, we show that 
with another suitable ansatz the static solutions of the sine-Gordon equation 
as well as the (time-dependent) solutions of the hyperbolic plane pendulum, 
inverted plane pendulum as well as the plane pendulum equation are all related 
to the solutions of a different nonlinear equation. We then present four 
solutions of this nonlinear equation and hence obtain four solutions of the 
above mentioned four equations. With yet another ansatz, we obtain a few new 
exact solutions of the plane pendulum problem. We then consider the plane 
pendulum problem in the first anharmonic approximation and show that the 
resulting equation is very similar to the equation for the static solutions of 
the symmetric $\phi^4$ 
model. Using this correspondence we obtain a large number of exact periodic 
and hyperbolic solutions of the plane pendulum problem in the first anharmonic 
approximation. For each periodic solution, we also mention the corresponding 
time period $T$. On the other hand, if we consider the plane pendulum in the 
second anharmonic approximation, we find that the resulting equation is very 
similar to the equation for the static solutions of the 
$\phi^2$-$\phi^4$-$\phi^6$ model but with the $\phi^4$ and the $\phi^6$ 
coefficients being highly constrained. As a result, so far we have been able to
 obtain only two exact solutions of the plane pendulum problem in the 
second anharmonic approximation which are presented in the Appendix. We also 
introduce the so called elliptic pendulum equation 
$\theta_{tt} = -\sn(\theta,m)/\dn(\theta,m)$ where $0 \le m \le 1$ is the 
modulus parameter while $\sn(\theta,m)$, $\dn(\theta, m)$, and $\cn(\theta,m)$ 
(see below) are the Jacobi elliptic functions \cite{as}. In the limit $m = 0$, 
the elliptic pendulum equation goes over to the plane pendulum equation while 
in the $m = 1$ limit it goes over to the hyperbolic plane pendulum equation. We
 show that in the harmonic approximation, no matter what is the value of 
$0 \le m \le 1$, the elliptic pendulum is an example of an isochronous system. 
The latter means that all solutions are periodic with the same fixed period so 
that the time period is independent of the amplitude. Further, for the special
case of the modulus parameter $m = 1/2$, it continues to be an isochronous 
system even in the first anharmonic approximation. For completeness,
we also consider the hyperbolic plane pendulum equation and obtain a few of
its solutions in the harmonic and the first anharmonic approximation. 

The plan of the paper is the following. In Sec. IIa we show that with a suitable
ansatz, the solutions of the static SG equation, static SHG equation and the 
plane pendulum as well as the inverted plane pendulum equations can all be 
shown to be related to the solution of a single nonlinear equation. We then 
obtain five exact solutions of this nonlinear equation (and hence exact 
solutions of all four problems in a single attempt). In Sec. 
IIb we show that with another suitable ansatz, the solutions of the static SG 
equation, the plane pendulum equation, the inverted pendulum equation and the 
hyperbolic pendulum equation can all be reduced to the solution of another 
nonlinear equation. We then obtain four exact solutions of this nonlinear 
equation (and hence the solutions of the four problems in a single attempt). 
In Sec. IIc we obtain a few more new solutions of the plane 
pendulum equation using another ansatz. In Sec. III we consider the plane 
pendulum equation in 
the first anharmonic approximation and point out its similarity to 
the static symmetric $\phi^4$ equation. Using this correspondence, we then 
obtain several exact periodic as well as hyperbolic solutions of the plane 
pendulum problem in the first anharmonic approximation. For each 
periodic solution we also mention the corresponding time period. 
In Sec. IV we introduce a novel elliptic pendulum equation which goes over 
to the plane pendulum and the hyperbolic plane pendulum equations in the 
appropriate limits. We analyze the elliptic pendulum equation in some detail 
and show that in the harmonic approximation it offers a one-parameter family of 
isochronous system. We also obtain exact solutions of the elliptic pendulum 
problem in the first anharmonic approximation and in the special case of $m = 
1/2$ we show that it continues to be an isochronous system even in the first 
anharmonic approximation. In Sec. V we summarize the results obtained in this 
paper and point out some of the open problems. 
In Appendix A we consider the plane pendulum equation in the second anharmonic 
approximation and point out its similarity with the static 
$\phi^2$-$\phi^4$-$\phi^6$ model with constraints on the $\phi^4$ and $\phi^6$ 
coefficients because 
of which we are able to obtain only two exact solutions. In Appendix B we 
obtain a few of the exact solutions of the hyperbolic plane pendulum equation 
in the harmonic and the first anharmonic approximation. Finally, in Appendix C 
we mention a few important properties of the Jacobi elliptic functions 
\cite{as} which are used in this paper.

\section{Connection Between the Solutions of the Plane Pendulum Equation 
with the Solutions of Several Important Equations in Physics}

One of the most important problems in dynamical systems is the plane pendulum,  
the equation of motion for which is given by
\be\label{1}
\theta_{tt} = - a\sin(\theta)\,,~~a = {g}/{l}\,.
\ee
By rescaling $t \rightarrow t/\sqrt{a}$ we can always eliminate $a$ from 
Eq. (\ref{1}) and write the plane pendulum equation as
\be\label{2}
\theta_{tt} = -\sin(\theta)\,.
\ee
Once we have obtained a solution $f(t)$ of the scaled Eq. (\ref{2}), it then 
follows that the corresponding solution of the original plane pendulum 
Eq. (\ref{1}) is $f(\sqrt{a}t)$. 

\subsection{Connection Between the Solutions of Plane Pendulum Equation,
Inverted Plane Pendulum Equation as well as Static Solutions of SG and 
SHG Equations}

We now show that the static solutions of the SG and the SHG equations and the 
(time dependent) solutions of the plane pendulum equation (\ref{2}) and the 
inverted plane pendulum equation (see below) are all intimately related 
to the solutions of the nonlinear equation
\be\label{3}
(1-u^2) u_{yy} + u (u_y)^2 = - u (1-u^2)^2\,,
\ee
where $y$ is $x$ or $t$ depending on the equation concerned. 
The proof is straightforward. We discuss the four cases one by one.

\noindent{\bf Case I: Sine-Gordon Equation}

Consider the sine-Gordon equation
\be\label{4}
\phi_{xx} = \sin[\phi(x)]\,.
\ee
On using the ansatz
\be\label{5}
u(x) = \cos[\phi(x)/2]\,,
\ee
in Eq. (\ref{4}) we find that in that case $u$ must satisfy the nonlinear 
equation essentially given by Eq. (\ref{3}), i.e.
\be\label{6}
(1-u^2) u_{xx} + u (u_x)^2 = - u (1-u^2)^2\,.
\ee

\noindent{\bf Case II: Sine Hyperbolic-Gordon Equation}

Consider the sine hyperbolic-Gordon equation
\be\label{7}
\phi_{xx} = \sinh[\phi(x)]\,.
\ee
On using the ansatz
\be\label{8}
u(x) = \cosh[\phi(x)/2]\,,
\ee
in Eq. (\ref{7}) we find that in that case $u$ must satisfy the same nonlinear 
equation, Eq. (\ref{6}). 

\noindent{\bf Case III: Plane Pendulum Equation}             

On using the ansatz
\be\label{9}
u(t) = \sin[\theta(t)/2]\,,
\ee
in the plane pendulum Eq. (\ref{2}) we find that in that case $u$ must satisfy 
essentially the same nonlinear Eq. (\ref{3}), i.e. the equation
\be\label{10}
(1-u^2) u_{tt} + u (u_t)^2 = - u (1-u^2)^2\,.
\ee

\noindent{\bf Case IV: Inverted Plane Pendulum Equation}             

Consider the inverted plane pendulum equation
\be\label{1a}
\theta_{tt} = +a \sin(\theta)\,,~~a = \frac{g}{l}\,.
\ee
As in the plane pendulum case, by rescaling $t \rightarrow t/\sqrt{a}$ we can 
always eliminate $a$ from Eq. (\ref{1a}) and write the inverted plane pendulum 
equation as
\be\label{2a}
\theta_{tt} = \sin(\theta)\,.
\ee
Thus, once we obtain a solution $f(t)$ of Eq. (\ref{2a}) then the corresponding 
solution of the inverted plane pendulum Eq. (\ref{1a}) is simply $f(\sqrt{a}t)$.
It is interesting to note that the inverted plane pendulum Eq. (\ref{2a}) has 
the same form as the static SG Eq. (\ref{4}) except $t$ is to be replaced 
by $x$ as we go from the inverted plane pendulum to the static SG equation.

On using the ansatz
\be\label{9a}
u(t) = \cos[\theta(t)/2]\,,
\ee
in the inverted plane pendulum Eq. (\ref{2}) we find that in that case $u$ must
satisfy the same nonlinear equation, Eq. (\ref{10}).

Before we discuss the solutions of Eq. (\ref{3}), it is worth noting that
Eq. (\ref{3}) is invariant under $(u,y) \rightarrow -(u,y)$ so that if $u_1(y)$
is a solution of Eq. (\ref{3}) then so is the solution $-u_1(y)$ as well as
$u_1(-y)$.
Notice that once we have a solution $u_1(y)$ of the nonlinear Eq. (\ref{3}), 
then the corresponding static solution of the SG Eq. (\ref{4}) is given by 
\be\label{11}
\phi_{SG}(x) = 2\cos^{-1}[u_1(x)]\,.
\ee
On the other hand, the corresponding static solution of the SHG Eq. (\ref{7}) 
is given by 
\be\label{12}
\phi_{SHG}(x) = 2\cosh^{-1}[u_1(x)]\,.
\ee
Further, the solution of the inverted plane pendulum Eq. (\ref{2a}) is given by 
\be\label{11a}
\theta_{invpen}(t) = 2\cos^{-1}[u_1(t)]\,.
\ee
Finally, the corresponding solution of the plane pendulum Eq. (\ref{2}) is 
given by 
\be\label{13}
\theta_{pen}(t) = 2\sin^{-1}[u_1(t)]\,.
\ee

We now present five solutions of Eq. (\ref{3}) so that one has immediately five
exact solutions of the static SHG Eq. (\ref{7}), plane pendulum Eq. (\ref{2}) 
(and hence Eq. (\ref{1})), inverted plane pendulum Eq. (\ref{2a}) (and hence 
Eq. (\ref{1a})) and static solutions of the SG Eq. (\ref{4}).

{\bf Solution I}

It is easy to check that Eq. (\ref{3}) admits the periodic kink solution 
\cite{as}
\be\label{14}
u(y) = \sn\left(\frac{y}{\sqrt{n}},n\right)\,,~~ 0 < n \le 1\,,
\ee
where $n$ is the modulus of the Jacobi elliptic functions. In Appendix C we 
have mentioned a few important properties of the Jacobi elliptic functions.

{\bf Solution II}

Another solution to Eq. (\ref{3}) is
\be\label{15}
u(y) = \sqrt{n}\sn(y,n)\,,~~0 < n \le 1\,.
\ee

{\bf Solution III}

In the limit $n = 1$, both the solutions I and II go over to the hyperbolic 
kink solution
\be\label{16}
u(y) = \tanh(y)\,,
\ee
which is the famous ``sticking" solution of the plane pendulum equation
\cite{cs,pk}. 

{\bf Solution IV}

Eq. (\ref{3}) also admits the solution
\be\label{17}
u(y) = \frac{\sqrt{n} \cn(y,n)}{\dn(y,n)}\,,~~0 < n < 1\,.
\ee

{\bf Solution V}

Remarkably, Eq. (\ref{3}) also admits a complex PT-invariant kink solution with
PT-eigenvalue $-1$
\be\label{18}
u(y) = \tanh(\beta y) + i\sech(\beta y)\,,
\ee
provided $\beta = 2$.

\subsection{Connection Between the Solutions of the Plane Pendulum with
those of the Inverted Plane Pendulum, the Hyperbolic Plane Pendulum and the 
Static Solutions of the SG Equation}
 
We now show that the exact solutions of the plane pendulum Eq. (\ref{2}), the
inverted plane pendulum Eq. (\ref{2a}), the hyperbolic plane pendulum (see 
below) and the static solutions of the SG Eq. (\ref{4}) are all intimately 
related to the solution of the nonlinear equation
\be\label{19}
(1-v^2) v_{yy} + v (v_y)^2 =  v (1-v^2)^2\,,
\ee
where $y$ is $x$ or $t$ depending on the equation concerned. 
The proof is straightforward. We discuss the four cases one by one. \\

\noindent{\bf Case I: Hyperbolic Plane Pendulum Equation}

Consider the hyperbolic plane pendulum equation
\be\label{20}
\theta_{tt} = -a\sinh(\theta)\,,~~a = \frac{g}{l}\,.
\ee
By rescaling $t \rightarrow t/\sqrt{a}$ we can always eliminate $a$ from 
Eq. (\ref{20}) and write the hyperbolic plane pendulum equation as
\be\label{21}
\theta_{tt} = - \sinh(\theta)\,.
\ee
Thus, once we obtain a solution $f(t)$ of Eq. (\ref{21}) then the corresponding 
solution of the hyperbolic plane pendulum Eq. (\ref{20}) is simply 
$f(\sqrt{a}t)$.

On using the ansatz
\be\label{22}
v(t) = \cosh[\theta(t)/2]\,,
\ee
in Eq. (\ref{21}) we find that in that case $v$ must satisfy essentially the 
same nonlinear Eq. (\ref{19}), i.e. 
\be\label{23}
(1-v^2) v_{tt} + v (v_t)^2 = v (1-v^2)^2\,.
\ee

\noindent{\bf Case II: Sine-Gordon Equation}

Consider the sine-Gordon Eq. (\ref{4}). On using the ansatz
\be\label{24}
v(x) = \sin[\phi(x)/2]\,,
\ee
in Eq. (\ref{4}) we find that in that case $v$ must satisfy the nonlinear 
equation essentially given by Eq. (\ref{19}), i.e.
\be\label{25}
(1-v^2) v_{xx} + v (v_x)^2 = v (1-v^2)^2\,.
\ee

\noindent{\bf Case III: Inverted Pendulum Equation}

Consider the inverted plane pendulum Eq. (\ref{2a}). On using the ansatz
\be\label{24a}
v(t) = \sin[\theta(t)/2]\,,
\ee
in Eq. (\ref{2a}) we find that in that case $v$ must satisfy the same nonlinear
Eq. (\ref{23}).

\noindent{\bf Case IV: Plane Pendulum Equation}             

On using the ansatz
\be\label{26}
v(t) = \cos[\theta(t)/2]\,,
\ee
in the plane pendulum Eq. (\ref{2}) we find that in that case $v$ must satisfy 
the same nonlinear Eq. (\ref{23}).
Before we discuss the solutions of Eq. (\ref{19}), it is worth noting that
Eq. (\ref{19}) is invariant under $(v,y) \rightarrow -(v,y)$ so that if 
$v_1(y)$ is a solution of Eq. (\ref{19}) then so is the solution $-v_1(y)$ as 
well as $v_1(-y)$.

Thus once we have a solution $v_1(y)$ of the nonlinear Eq. (\ref{19}), 
then the corresponding static solution of the SG Eq. (\ref{4}) is given by 
\be\label{27}
\phi_{SG}(x) = 2\sin^{-1}[v_1(x)]\,.
\ee
On the other hand, the corresponding solution of the hyperbolic pendulum 
Eq. (\ref{21}) is given by 
\be\label{28}
\theta_{hpen}(t) = 2\cosh^{-1}[v_1(t)]\,.
\ee
Further, the corresponding solution of the inverted plane pendulum  
Eq. (\ref{2a}) is given by 
\be\label{27b}
\theta_{invpen}(t) = 2\sin^{-1}[v_1(t)]\,.
\ee
Finally, the corresponding solution of the plane pendulum Eq. (\ref{2}) is 
given by 
\be\label{29}
\theta_{Pen}(t) = 2\cos^{-1}[v_1(t]\,.
\ee

We now present four solutions of Eq. (\ref{19}) so that one has immediately 
four exact solutions of the hyperbolic plane pendulum Eq. (\ref{21}) (and hence
Eq. (\ref{20})), the plane pendulum Eq. (\ref{2}) (and hence Eq. (\ref{1})), the 
inverted plane pendulum Eq. (\ref{2a}) (and hence Eq. (\ref{1a})) and the 
static solutions of the SG Eq. (\ref{4}).

{\bf Solution I}

It is easy to check that Eq. (\ref{19}) admits the solution
\be\label{30}
u(y) = \cn\left(\frac{y}{\sqrt{n}},n\right)\,.
\ee

{\bf Solution II}

Another solution to Eq. (\ref{19}) is
\be\label{30a}
u(y) = \dn(y,n)\,.
\ee

{\bf Solution III}

In the limit $n = 1$, both the solutions go over to the hyperbolic solution
\be\label{31}
u(y) = \sech(y)\,.
\ee

{\bf Solution IV}

Eq. (\ref{19}) also admits the solution
\be\label{27a}
u(y) = \frac{\sqrt{1-n}}{\dn(y,n)}\,,~~0 < n < 1\,.
\ee
Unfortunately, with the second ansatz, we have not been able to obtain an 
analogue of the complex PT-invariant solution V as given by Eq. (\ref{18}) 
using the 
first ansatz. In the case of the plane pendulum, the inverted plane pendulum as
well as the SG case, it is easily shown that the first four solutions with the 
first and the second ansatz are one and the same solution. 

\subsection{Few More Exact Solutions of the Plane Pendulum Equation}

We now obtain a few more exact solutions of the plane pendulum Eq. (\ref{1}) by
using the ansatz
\be\label{32}
\tan[\phi(t)/4] = w\,.
\ee
It turns out that these are new solutions distinct from those presented in
Secs. IIa and IIb above.
On using the ansatz in the pendulum Eq. (\ref{1}), one finds that $w$ must
satisfy the nonlinear equation
\be\label{33}
(1+w^2) w_{tt} -2 w(w_t)^2 = -aw (1-w^2)\,,~~a > 0\,.
\ee
Before we discuss the solutions of Eq. (\ref{33}), it is worth noting that
Eq. (\ref{33}) is invariant under $(w,t) \rightarrow -(w,t)$ so that if 
$w_1(t)$ is a solution of Eq. (\ref{33}) then so is the solution $-w_1(t)$ as 
well as $w_1(-t)$. We now show that the nonlinear Eq. (\ref{33}) admits the 
following five exact solutions.

{\bf Solution I}

It is easy to check that the nonlinear Eq. (\ref{33}) (and hence the plane 
pendulum  Eq. (\ref{1})) admit the real periodic solution
\be\label{34}
w(t) = A \sn(\beta t,n)\,,
\ee
provided
\be\label{35}
A = \pm (n)^{1/4}\,,~~a = (1+\sqrt{n})^2 \beta^2\,.
\ee
Note that this solution holds good in case $0 < n \le 1$. In particular, in
case $n = 1$ we obtain the hyperbolic solution
\be\label{36}
w(t) = A \tanh(\beta t)\,,
\ee
provided
\be\label{37}
A = 1\,,~~a = 4 \beta^2\,.
\ee

{\bf Solution II}

Eq. (\ref{33}) also admits an imaginary solution
\be\label{38}
w(t) = iA \sn(\beta t,n)\,,
\ee
provided
\be\label{39}
A = \pm (n)^{1/4}\,,~~a = (1-\sqrt{n})^2 \beta^2\,.
\ee
Note that this solution holds good for $0 < n < 1$.

{\bf Solution III}

Even though neither $\cn(t,n)$ nor i$\cn(t,n)$ is a solution of Eq. (\ref{33}), it 
turns out that Eq. (\ref{33}) admits the complex superposed PT-invariant 
solution with PT-eigenvalue $-1$ 
\be\label{40}
w(t) = A[\sn(\beta t,n)\pm i\cn(\beta t,n)]\,,
\ee
provided
\be\label{41}
A^2 = 1\,,~~a = \beta^2\,.
\ee

{\bf Solution IV}

Even though neither $\cn(t,n)$ nor i$\cn(t,n)$ is a solution of Eq. (\ref{33}),
it turns out that Eq. (\ref{33}) also admits the complex superposed 
PT-invariant solution with PT-eigenvalue $+1$
\be\label{42}
w(t) = A[\cn(\beta t,n)\pm i\sn(\beta t,n)]\,,
\ee
provided
\be\label{43}
A^2 = 1\,,~~a = (1-n)\beta^2\,.
\ee

{\bf Solution V}

Remarkably, even though neither $\dn(\beta t,m)$ nor $i\dn(\beta t,m)$ 
is a solution of Eq. (\ref{33}), it turns out that Eq. (\ref{33}) admits the 
complex PT-invariant periodic superposed solution with PT-eigenvalue $-1$
\be\label{44}
w(t) = A[\sqrt{n}\sn(\beta t,n)\pm i \dn(\beta t,n)]\,,
\ee
provided
\be\label{45}
A^2 = 1\,,~~a = n \beta^2\,.
\ee
Note that the solution holds good only if $0 < n \le 1$. In particular, in the
limit $n =1$, both solutions III and V go over to the complex PT-invariant
(hyperbolic) kink solution with PT-eigenvalue $-1$
\be\label{46}
w(t) = A[\tanh(\beta t)\pm i \sech(\beta t)]\,,
\ee
provided
\be\label{47}
A^2 = 1\,,~~a = \beta^2\,.
\ee

It is easy to check that these five solutions are distinct from the solutions 
of the pendulum Eq. (\ref{1}) presented in Secs. IIa and IIb.

\section{Periodic and Hyperbolic Solutions of the Plane Pendulum 
Equation in the First Anharmonic Approximation}

We shall now present several periodic as well as hyperbolic solutions of the
plane pendulum in the first anharmonic approximation. In the case of the 
periodic solutions we also explicitly mention the corresponding period. 

We start from the plane pendulum Eq. (\ref{1}). As is well known, in the 
harmonic approximation, i.e. the case of small $\theta$, the pendulum
Eq. (\ref{1}) is given by
\be\label{1.1}
\theta_{tt} = -a \theta\,,~~a = \frac{g}{l}\,,
\ee
whose solution is 
\be\label{1.2}
\theta(t) = A\sin(\sqrt{a}t)\,,
\ee
with period $T = 2\pi \sqrt{{l}/{g}}$. As is well known, for this solution 
the time period is independent of the amplitude, which is the best well 
known example of an isochronous system.

In the first anharmonic approximation, we have $\sin(\theta) = \theta 
-\frac{\theta^3}{6}+...$, so that Eq. (\ref{1}) takes the form
\be\label{1.3}
\theta_{tt} = -a\theta + \frac{a\theta^3}{6}-...\,.
\ee

We shall now obtain several exact solutions of the pendulum Eq. (\ref{1.3}). 
It is worth pointing out that Eq. (\ref{1.3}) is analogous to the symmetric
$\phi^4$ equation in the case of the static solutions
\be\label{1.4}
\phi_{xx} = p x + q x^3\,, 
\ee
for which a large number of solutions are already known. Thus using the analogy
and replacing $x$ with $t$, we can immediately write down a large number of 
exact periodic and hyperbolic solutions of the pendulum Eq. (\ref{1.3}). On
comparing Eqs. (\ref{1.3}) and (\ref{1.4}), it is clear that in the pendulum
case $p = -a < 0\,,~~q = \frac{a}{6} > 0$. Note that $a = \frac{g}{l}$. Using
this $\phi^4$ analogy we now present 12 solutions of Eq. (\ref{1.3}). For 
clarity we shall first present solutions of a more general equation
\be\label{1.5}
\theta_{tt} = -a\theta + b\theta^3\,,~~a, b > 0\,,
\ee
and then find out the conditions under which these are solutions of the 
plane pendulum Eq. (\ref{1.3}), i.e. when $b = a/6$. 

{\bf Solution I}

Eq. (\ref{1.5}) admits  \cite{aubry, CKMS} the periodic kink solution
\be\label{1.5a}
\theta(t) = A\sqrt{n} \sn(\beta t,n)\,,
\ee
provided
\be\label{1.6}
b A^2 = 2 \beta^2\,,~~a = (1+n)\beta^2\,. 
\ee
Thus, it is a solution of the plane pendulum Eq. (\ref{1.3}) in case
\be\label{1.7}
A = \frac{2\sqrt{3}}{\sqrt{1+n}}\,,
\ee
and the corresponding time period is
\be\label{1.8a}
T(n) = 4K(n)\sqrt{\frac{12}{a A^2}}\,,
\ee
where $K(n)$ is the complete elliptic integral of the first kind \cite{as}. 
Note that since $a = g/l$, hence dimensionally we get the correct equation for
the time period and one can see how far different it is from the lowest order
expression. Further, note that unlike the harmonic approximation case, now 
the time period depends on the amplitude (i.e. $T \propto A^{-1}$), i.e. it is 
not an isochronous system in the first anharmonic approximation. It turns out 
that this is the case for all the periodic solutions of the plane pendulum. 
Thus, while all of them have correct dimensional dependence but the time period 
$T \propto A^{-1}$  for all the solutions mentioned below. We will therefore 
not mention it again while discussing other periodic solutions of the 
pendulum Eq. (\ref{1.3}).

{\bf Solution II}

In the limit $m = 1$ the solution I goes over to the (hyperbolic) kink solution
\be\label{1.9}
\theta(t) = A \tanh(\beta t)\,,
\ee
provided
\be\label{1.10}
b A^2 = 2 \beta^2\,,~~a = 2\beta^2\,. 
\ee
Thus as $t$ goes from $-\infty$ to $\infty$, $x(t)$ goes from $-A$ to $A$ 
where $A = \sqrt{6}$.

{\bf Solution III}\\
Eq. (\ref{1.5})  also admits a complex
PT-invariant kink solution with $PT$ eigenvalue $-1$ \cite{ks1}
\be\label{1.11}
\theta(t) = \sqrt{n}[A \sn(\beta t, n) +i B \cn(\beta t, n)]\,,
\ee
provided 
\be\label{1.12}
B = \pm A\,,~~2bA^2 =  \beta^2\,,~~a = \frac{(2-n)\beta^2}{2}\,.
\ee
Thus, it is a solution of the plane pendulum Eq. (\ref{1.3}) in case
\be\label{1.13}
A = \frac{\sqrt{6}}{\sqrt{2-n}}\,,
\ee
and the corresponding time period is
\be\label{1.14}
T(n) = 4K(n)\sqrt{\frac{3}{a A^2}}\,.
\ee

{\bf Solution IV}

Remarkably, Eq. (\ref{1.5}) also admits another complex PT-invariant periodic 
kink solution with $PT$ eigenvalue $-1$
\be\label{1.15}
\theta(t) = A \sqrt{n} \sn(\beta t, n) +i B \dn(\beta t, n)\,,
\ee
provided $1/2 < n < 1$ and further 
\be\label{1.16}
B = \pm A\,,~~2bA^2 =  \beta^2\,,~~a = \frac{(2n-1)\beta^2}{2}\,.
\ee
Thus, it is a solution of the plane pendulum Eq. (\ref{1.3}) in case
\be\label{1.17}
A = \sqrt{\frac{6}{(2n-1)}}\,,
\ee
and the corresponding time period is
\be\label{1.18}
T(n) = 4K(n)\sqrt{\frac{3}{a A^2}}\,.
\ee

{\bf Solution V}\\
In the limit $n = 1$, both solutions III and IV go over to the
PT-invariant (hyperbolic) complex kink solution with $PT$ eigenvalue $-1$ 
\cite{ks1}
\be\label{1.19}
\theta(t) = [A \tanh(\beta t) +i B \sech(\beta t)]\,,
\ee
provided 
\be\label{1.20}
B = \pm A\,,~~2bA^2 =  \beta^2\,,~~a = \frac{\beta^2}{2}\,,
\ee
where $A = \sqrt{6}$.

{\bf Solution VI} \\
Eq. (\ref{1.5}) admits the periodic solution \cite{ks4}
\be\label{1.21}
\theta(t) = \frac{A \dn(\beta t,n) \cn(\beta t,n)}{1+B\cn^2(\beta t,n)}\,,
~~B > 0\,,
\ee
provided
\bea\label{1.22}
&&0 < n < 1\,,~~B = \frac{\sqrt{n}}{1-\sqrt{n}}\,,
~~a = [1+n+6\sqrt{n}]\beta^2\,, \nonumber \\
&&b A^2 = \frac{8 \sqrt{n} \beta^2}{(1-\sqrt{n})^2}\,.
\eea
Thus, Eq. (\ref{1.21}) is a solution of the plane pendulum Eq. (\ref{1.3}) in case
\be\label{1.23}
A = \frac{4\sqrt{3\sqrt{n}}}{(1-\sqrt{n})\sqrt{(1+n+6\sqrt{n})}}\,,
\ee
and the corresponding time period is
\be\label{1.24}
T(n) = 4K(n)\sqrt{\frac{48m}{(1-\sqrt{n})^2 a A^2}}\,.
\ee

On using the identity \cite{as}
\be\label{1.25}
\frac{\cn(y,n)\dn(y,n)}{1+B\cn^2(y,n)} = \frac{\dn^2(\Delta,n)}
{2\sn(\Delta,n)}[\sn(y+\Delta,n)-\sn(y-\Delta,n)]\,,
\ee
where $B = {n\sn^2(\Delta,n)}/{\dn^2(\Delta,n)}$, one can re-express 
the solution (\ref{1.25}) as
\be\label{1.26}
\theta(t) = \frac{\sqrt{2n}\beta}{|b|}[\sn(\beta t+\Delta,n)
-\sn(\beta t -\Delta,n)]\,,
\ee
where $\Delta$ is defined by $\sn(\sqrt{n}\Delta,1/n) = \pm n^{1/4}$,
and use has been made of the identity \cite{as} 
\be\label{1.27}
\sqrt{n} \sn(y,n) = \sn(\sqrt{n} y,1/n)\,. 
\ee

{\bf Solution VII} \\
It is easy to check that \cite{ks3}
\be\label{1.28}
\theta(t) = \frac{A\sqrt{n}\sn(\beta t,n)}{D+\dn(\beta t,n)}\,,
\ee
is an exact periodic pulse solution of Eq. (\ref{1.5}) provided
\be\label{1.29}
D = 1\,,~~2 b A^2 =  \beta^2\,,~~a = (2-n)\frac{\beta^2}{2}\,.
\ee
Thus, it is a solution of the plane pendulum Eq. (\ref{1.3}) in case
\be\label{1.30}
A = \frac{\sqrt{6}}{\sqrt{2-n}}\,,
\ee
and the corresponding time period is
\be\label{1.31}
T(n) = 4K(n)\sqrt{\frac{3}{a A^2}}\,.
\ee

{\bf Solution VIII}

One finds that \cite{ks3} 
\be\label{1.32}
\theta(t) = \frac{A\sqrt{n}\cn(\beta t,n)}{D+\dn(\beta t,n)}\,,
\ee
is an exact periodic pulse solution of Eq. (\ref{1.5}) 
provided
\be\label{1.33}
0 < n < 1\,,~~D^2 = 1-n \,,~~2 b A^2 =  \beta^2\,,~~a 
= (2-n)\frac{\beta^2}{2}\,.
\ee
Thus, it is a solution of the plane pendulum Eq. (\ref{1.3}) in case
\be\label{1.34}
A = \frac{\sqrt{6}}{\sqrt{(2-n)}}\,,
\ee
and the corresponding time period is
\be\label{1.35}
T(n) = 4K(n)\sqrt{\frac{3}{a A^2}}\,.
\ee

{\bf Solution IX}

We find that  that \cite{ks3} Eq. (\ref{1.5}) also admits a 
complex PT-invariant periodic superposed pulse solution with 
PT-eigenvalue +$1$ 
\be\label{1.36}
\theta(t) = \frac{\sqrt{n}[A\cn(\beta t,n)+iB\sn(\beta t,n)]}
{D+\dn(\beta t,n)}\,,~~D^2 > 1\,,
\ee
provided $0 < n <1/2$ and further
\be\label{1.37}
2 b A^2 = (D^2-1)\beta^2\,,~~2 b B^2 = (D^2-1+n)\beta^2\,,~~
a = (1-2n)\frac{\beta^2}{2}\,.
\ee
Thus, it is a solution of the plane pendulum Eq. (\ref{1.3}) in case
\be\label{1.38}
A = \frac{\sqrt{6(D^2-1)}}{\sqrt{(1-2n)}}\,. 
\ee
This is a periodic solution with time period $T$ being 
\be\label{1.39}
T(n) = 4K(n)\sqrt{\frac{3(D^2-1)}{a A^2}}\,. 
\ee

{\bf Solution X}

Eq. (\ref{1.5}) also admits a complex PT-invariant periodic 
kink superposed solution with PT-eigenvalue $-1$ \cite{ks3}
\be\label{1.40}
\theta(t) = \frac{\sqrt{n}[A\sn(\beta t,n)+iB\cn(\beta t,n)]}
{D+\dn(\beta t,n)}\,,~~ D^2 > 1\,,
\ee
provided
\be\label{1.41}
2 b n A^2 = (D^2-1+n)\beta^2\,,~~2 b B^2 = (D^2-1)\beta^2\,,~~
a = (2-n)\frac{\beta^2}{2}\,.
\ee
Thus, it is a solution of the plane pendulum Eq. (\ref{1.3}) in case
\be\label{1.42}
A = \frac{\sqrt{6(D^2-1+n)}}{\sqrt{n(2-n)}}\,.
\ee
This is a periodic solution with time period $T$ being 
\be\label{1.43}
T(n) = 4K(n)\sqrt{\frac{3(D^2-1+n)}{a A^2}}\,. 
\ee

{\bf Solution XI}

Remarkably, even a real superposition  
\be\label{1.44}
\theta(t) = \frac{\sqrt{n}[A\sn(\beta t,n)+B\cn(\beta t,n)]}
{D+\dn(\beta t,n)}\,,~~D > 0\,,
\ee
is an exact periodic superposed mixed parity solution (i.e. an admixture of 
periodic kink and periodic pulse solutions) of Eq. (\ref{1.5}) provided
\be\label{1.45}
2 b A^2 = (D^2-1+n) \beta^2\,,~~2b B^2 = (1-D^2) \beta^2\,,~~
a = (2-n)\frac{\beta^2}{2}\,.
\ee
Thus, this solution exists only if $(1-n) < D^2 < 1$, $b > 0$, $a < 0$ and 
$0 < n \le 1$. 
It is a solution of the plane pendulum Eq. (\ref{1.3}) in case
\be\label{1.46}
A = \frac{\sqrt{6(D^2-1+n)}}{\sqrt{(2-n)}}\,. 
\ee
This is a periodic solution with time period $T$ being 
\be\label{1.47}
T(n) = 4K(n)\sqrt{\frac{3(D^2-1+n)}{a A^2}}\,. 
\ee

{\bf Solution XII}

We also find that \cite{ks3}
\be\label{1.48}
\theta(t) = \frac{[A\sqrt{n}\sn(\beta t,n)+iB\dn(\beta t,n)]}
{D+\cn(\beta t,n)}\,,~~D > 1\,,
\ee
is an exact complex PT-invariant periodic solution with PT-eigenvalue 
$-1$ of Eq. (\ref{1.5}) provided $1/2 < n < 1$ and further
\be\label{1.49}
2 n b A^2 = (1-n+n D^2)\beta^2\,,~~2 b B^2 = (D^2-1)\beta^2\,,~~
a = (2n-1)\frac{\beta^2}{2}\,.
\ee
Thus, it is a solution of the plane pendulum Eq. (\ref{1.3}) in case
\be\label{1.50}
A = \frac{\sqrt{6(nD^2+1-n)}}{\sqrt{n(2n-1)}}\,. 
\ee
This is a periodic solution with time period $T$ being 
\be\label{1.51}
T(n) = 4K(n)\sqrt{\frac{3(1-n+nD^2)}{a n A^2}}\,. 
\ee

\section{Elliptic Pendulum}

It is worth asking if one can generalize the plane pendulum Eq. (\ref{1}) and 
introduce a one-parameter family of so called ``elliptic pendulum" equations
which will reduce to the plane pendulum Eq. (\ref{1}) in one limit and 
hyperbolic pendulum Eq. (\ref{20}) in the other limit. The answer to the 
question is in the affirmative. In particular, consider the elliptic plane 
pendulum equation
\be\label{2.1}
\theta_{tt} = -a\frac{\sn(\theta,m)}{\dn(\theta,m)}\,,
\ee
where $m$ is the modulus parameter $0 \le m \le 1$ \cite{as}. On using the fact
that
\bea\label{2.2}
&&\sn(\theta,m=0) = \sin(\theta)\,,~~\dn(\theta,m=0) = 1\,,~~\sn(\theta,m=1)
= \tanh(\theta)\,, \nonumber \\
&&\dn(\theta,m=1) = \sech(\theta)\,,
\eea
it immediately follows that while in the $m = 0$ limit, the elliptic pendulum
Eq. (\ref{2.1}) reduces to the plane pendulum Eq. (\ref{1}), in the $m = 1$
limit, it reduces to the hyperbolic pendulum Eq. (\ref{20}). So far we have not
been able to obtain exact analytical solutions of the elliptic pendulum 
Eq. (\ref{2.1}). However, it may be of interest to obtain solutions of the 
elliptic pendulum in the harmonic and the first anharmonic approximation.

\subsection{Periodic as well as Hyperbolic Solutions of the Elliptic Plane 
Pendulum Equation in the Harmonic and the First Anharmonic Approximation}

We start from the elliptic plane pendulum Eq. (\ref{2.1}). As is well known, in 
the case of small $\theta, \sn(\theta,m)$ and $\dn(\theta,m)$ to 
$O(\theta^5)$ are given by \cite{as}
\bea\label{2.3}
&&\sn(\theta,m) = \theta -\frac{(1+m)}{6} \theta^3 
+ \frac{(1+14m+m^2)\theta^5}{120}+...\,, \nonumber \\
&&\dn(\theta,m) = 1 - \frac{m}{2} \theta^2 + \frac{m(4+m) \theta^4}{24}
+ O(\theta^6)\,.
\eea
On substituting Eq. (\ref{2.3}) in Eq. (\ref{2.1}) and collecting
terms to $O(\theta^5)$ gives 
\be\label{2.4}
\theta_{tt} = -a\theta +\frac{a(1-2m)\theta^3}{6} 
-\frac{(1-16m+16m^2) a\theta^5}{120}\,.
\ee
From here we find that no matter what $0 \le m \le 1$ is, in the harmonic 
approximation the elliptic pendulum behaves as an isochronous system and
the solution is given by
\be\label{2.5}
\theta(t) = A\sin[\sqrt{a}(t+t_0)]\,,
\ee
with the time period 
\be\label{2.6}
T = \frac{2\pi}{\sqrt{a}} = 2\pi \sqrt{\frac{l}{g}}\,,
\ee
being independent of the amplitude $A$. Thus to $O(\theta)$, i.e. in the 
harmonic approximation, we have one continuous-parameter family of isochronous 
systems characterized by the modulus parameter $0 \le m \le 1$. Remarkably, 
for $m = 1/2$, even to $O(\theta^3)$ the solution of Eq. (\ref{2.4}) 
continues to be Eq. (\ref{2.5}) while the time period $T$ as given by Eq. 
(\ref{2.6}) continues to be independent of the amplitude $A$. Thus for 
$m = 1/2$ we have an isochronous system to $O(\theta^3)$. \\

\subsection{Exact Solutions of the Elliptic Plane Pendulum in the First 
Anharmonic Approximation in Case $m < 1/2$}

From Eq. (\ref{2.4}) it follows that for $m < 1/2$, in the first anharmonic 
approximation the elliptic plane pendulum equation is given by 
\be\label{2.7}
\theta_{tt} = -a\theta +\frac{a(1-2m)\theta^3}{6}\,.
\ee
On comparing it with the plane pendulum Eq. (\ref{1.3}) in the first anharmonic 
approximation, we conclude that for $m < 1/2$, the elliptic pendulum 
in the first anharmonic approximation essentially behaves like a plane 
pendulum except the $\theta^3$ coefficient $1/6$ in the plane pendulum case
gets replaced by $(1-2m)/6$. Thus, all the solutions of the plane pendulum
Eq. (\ref{1.3}) in the first anharmonic approximation are also the solutions
of the elliptic pendulum Eq. (\ref{2.7}) with the obvious replacement of
$1/6$ by $(1-2m)/6$ ($m < 1/2$) in all the solutions of the plane pendulum 
Eq. (\ref{1.3}) as given in Sec. III. Effectively, it means to replace the 
amplitude $A$ everywhere by $\sqrt{1-2m} A$ ($m < 1/2$) in all the solutions I 
to XII of the plane pendulum Eq. (\ref{1.3}). As an illustration, consider the 
solution I of  Eq. (\ref{1.3}) and discuss it in the context of the 
solution of the elliptic pendulum Eq. (\ref{2.7}). It is easy to check that 
Eq. (\ref{2.7}) admits the periodic kink solution
\be\label{2.8}
\theta(t) = A\sqrt{n} \sn(\beta t,n)\,,
\ee
provided
\be\label{2.9}
A\sqrt{1-2m} = \frac{2\sqrt{3}}{\sqrt{1+n}}\,,
\ee
and the corresponding time period is
\be\label{2.10}
T(n) = 4K(n)\sqrt{\frac{12}{a(1-2m) A^2}}\,.
\ee

\subsection{Exact Solutions of the Elliptic Plane Pendulum in the First 
Anharmonic Approximation in Case $m > 1/2$}

From Eq. (\ref{2.4}) it follows that for $m > 1/2$, in the first anharmonic 
approximation the elliptic plane pendulum equation is given by 
\be\label{2.11}
\theta_{tt} = -a\theta -\frac{a(2m-1)\theta^3}{6}\,.
\ee
On comparing it with the hyperbolic plane pendulum Eq. (\ref{19}) in the first 
anharmonic approximation, we conclude that for $m > 1/2$, the elliptic pendulum 
in the first anharmonic approximation essentially behaves like a hyperbolic 
plane pendulum except the $\theta^3$ coefficient $1/6$ in the hyperbolic 
plane pendulum case gets replaced by $(2m-1)/6$ ($m > 1/2$). Thus, all the 
solutions of the hyperbolic plane pendulum equation in the first anharmonic 
approximation (see Appendix B below) are also the solutions of the elliptic 
pendulum Eq. (\ref{2.11}) with the obvious replacement of $1/6$ by $(2m-1)/6$ 
($m > 1/2$) in all the solutions of the hyperbolic plane pendulum 
Eq. (\ref{20}) as given in Sec. III. This effectively means replacing amplitude
$A$ by $\sqrt{2m-1} A$ in all the 13 solutions of the hyperbolic plane pendulum
Eq. (\ref{B1}) as discussed in Appendix B.
 
\section{Summary and Open Problems}

In this paper we have pointed out a novel connection between the solutions of  
the plane pendulum problem and several different models. In particular, we have 
shown that the solutions of the plane pendulum, the inverted plane pendulum  
and the static solutions of the SG and the SHG equations are all related to the
solutions of the nonlinear Eq. (\ref{3}). We have also shown that the solutions
of the plane pendulum, the inverted plane pendulum, the hyperbolic plane 
pendulum and the static solutions of the SG equation are related to the 
solutions of the nonlinear Eq. (\ref{19}). Besides, we have obtained several 
exact periodic and hyperbolic solutions of the plane pendulum equation in the 
first anharmonic approximation. Finally, in Sec. IV we have introduced one 
continuous-parameter family of elliptic potentials which smoothly go over to 
the plane pendulum in the limit of the modulus parameter $m$ going to zero and 
to the hyperbolic plane pendulum in the limit of $m = 1$ and further shown that
in the harmonic approximation, the continuous parameter family of the elliptic 
plane pendulums (i.e. $0 \le m \le 1$) are examples of an isochronous system. 
Further, we have shown that for the special case of the modulus parameter 
$m = 1/2$, the elliptic pendulum is an isochronous system even up to the first 
anharmonic approximation (i.e., to $O(\theta^3)$). In 
Appendix B we have also obtained exact solutions of the hyperbolic plane 
pendulum equation in the harmonic as well as the first anharmonic approximation. 
This paper raises several open questions some of which are the following.

\begin{enumerate}

\item In Sec. IIa we have obtained five exact solutions of the nonlinear 
Eq. (\ref{3}). It is hard to believe that there are no other exact solutions
of this nonlinear equation. Clearly, finding even one more solution of 
Eq. (\ref{3}) would give us an additional exact solution of four different 
models, i.e. static solution of the SG and the SHG as well as (time dependent) 
solutions of the inverted plane pendulum and the plane pendulum equations. 

\item In Sec. IIb we have obtained four exact solutions of the nonlinear 
Eq. (\ref{19}). Again, it is hard to believe that there are no other exact 
solutions of this nonlinear equation. Clearly, finding even one more solution 
of Eq. (\ref{19}) would give us an additional exact solution of four different 
models, i.e. static solution of the SG as well as (time dependent) solutions of
the inverted, the hyperbolic and the plane pendulum equations. 

\item Turning around, if one can find even one new solution of any of the above
models, that in turn will also give us one extra solution to the three other 
interesting models. In particular, the SG equation has been examined for more 
than six decades now and it is hard to believe that it only admits five static 
solutions. Knowing one more new static solution of the SG equation will give us
a new solution of the plane, the hyperbolic and the inverted pendulum equations
as well as a static solution of the SHG equation.

\item In Sec. IIc we have obtained five new solutions of the nonlinear 
Eq. (\ref{33}) and hence the plane pendulum equation (\ref{1}). Again it is 
hard to believe that the nonlinear Eq. (\ref{33}) only admits five solutions 
and it is worth trying to obtain more exact solutions of the nonlinear 
Eq. (\ref{33}).

\item In Secs. III and Appendix A we have obtained several new solutions of the
plane pendulum equation in the first and the second anharmonic approximation 
respectively. It would be interesting if one can find physical significance of 
some of these solutions.

\item In Sec. IV we have introduced a one-parameter family of elliptic 
potentials and have unearthed some novel features of these elliptic pendulums.
Clearly, such an elliptic pendulum equation is not unique and one can come up 
with some other elliptic pendulum equation, one example being the one suggested
by Bender et al. \cite{ben}. It would be worthwhile examining the various 
possible choices and look for their unique features. In particular, can one 
think of an elliptic pendulum equation which at least for some modulus 
parameter $0 \le m \le 1$ is isochronous up to second or even higher order 
anharmonic approximation?

\end{enumerate}

\section{Appendix A: Solutions of the Plane Pendulum Equation in the Second 
Anharmonic Approximation}

We shall now present solutions of the plane pendulum equation in the second 
anharmonic approximation. As we show below, since the coefficients of $\theta$,
$\theta^3$ as well as $\theta^5$ are fixed and related to each other, this 
severely restricts the number of possible solutions. 

In the second anharmonic approximation, we have $\sin(\theta) = \theta 
-\frac{\theta^3}{6}+ \frac{\theta^5}{120}+...$, so that in this approximation
the plane pendulum Eq. (\ref{1}) takes the form
\be\label{A1}
\theta_{tt} = -a\theta + \frac{a\theta^3}{6} -\frac{a\theta^5}{120}\,.
\ee
This equation is of the form
\be\label{A2}
\theta_{tt} = -a\theta + b\theta^3 - d\theta^5\,,~~a,b,d > 0\,,
\ee
where
\be\label{A3}
b = \frac{a}{6}\,,~~\frac{b^2}{ad} = \frac{10}{3}\,.
\ee
This restriction of $a,b,d > 0$ and $b^2/ad = 10/3$ rules out most of the 
allowed solutions of Eq. (\ref{A2}). In fact we were able to obtain only 
two solutions of Eq. (\ref{A2}) with these restrictions which we now discuss 
one by one. 

{\bf Solution I}

It is easy to check that 
\be\label{A4}
\theta(t) = A\sqrt{\frac{[1-\dn(\beta t,n)]}{D+\dn(\beta t,n)}}\,,~~D > 0\,,
\ee
is an exact solution of Eq. (\ref{A2}) provided
\bea\label{A5}
&&(D-1)a = [(4+n)D+4-5n]\beta^2\,, \nonumber \\
&&(D-1)bA^2 = [D(2D-n)-2(1-n)]\beta^2\,, \nonumber \\
&&(D-1)d A^4 =\frac{3}{4}(D-1)[D^2+(1-n)]\beta^2\,.
\eea 
Since for our case $a,b,d > 0$, this implies that such a solution is possible 
only if $D > 1$. On demanding the restriction ${b^2}/{ad} = {10}/{3}$, 
we obtain a quartic equation for $D$ in terms of $m$ 
\bea\label{A6}
&&(12-5n)D^4 -2n D^3+(13n^2+54n-64)D^2+62n(1-n)D \nonumber \\
&&+(52-57n)(1-n) = 0\,.
\eea
We have explicitly checked that for $D \ge 3$ there is no acceptable solution
to this equation while there are acceptable solutions in case $1 < D \le 2$.
Thus one obtains a continuous-parameter family of solutions. On demanding
$b = a/6$, we find that $A^2$ is fixed in terms of $D$ and $m$, and is given by
\be\label{A7}
A^2 = \frac{6[2D^2-Dn-2(1-n)]}{(4+n)D+4-5n}\,.
\ee
As an illustration, when $D = 2$ we find that the nonlinear Eq. (\ref{A6}) is
satisfied if $n = \frac{27-\sqrt{681}}{6}$ which is approximately $n = 1/6$.
And in this case $A^2 = 3.17$.

{\bf Solution II}

It is easy to check that 
\be\label{A8}
\theta(t) = A\sqrt{\frac{[1-\cn(\beta t,n)]}{D+\cn(\beta t,n)}}\,,~~D > 1\,,
\ee
is an exact solution of Eq. (\ref{A2}) provided
\bea\label{A9}
&&(D+1)a = [(4n+1)D+4n-5]\beta^2\,, \nonumber \\
&&(D+1)bA^2 = [D(2n D-1)+2(1-n)]\beta^2\,, \nonumber \\
&&(D+1)d A^4 =\frac{3}{4}(D-1)[nD^2+(1-n)]\beta^2\,.
\eea 
On demanding the restriction ${b^2}/{ad} = {10}/{3}$, we obtain a 
quartic equation for $D$ in terms of $m$ 
\be\label{A10}
(12n^2-5n)D^4 -2n D^3+(8+4n-24n^2)D^2-2(1-n)D+(7-12n)(1-n) = 0\,.
\ee
We have explicitly checked that for $D \le 3$ there is no acceptable solution
to this equation while there are acceptable solutions in case $D \ge 4$.
Thus one obtains a continuous-parameter family of solutions. On demanding
$b = a/6$, we find that $A^2$ is fixed in terms of $D$ and $m$, and is given by
\be\label{A11}
A^2 = \frac{6[2n D^2-D+2(1-n)]}{(4+n)D+4n-5}\,.
\ee
As an illustration, when $D = 2$ we find that the Eq. (\ref{A10}) is
satisfied if $n = \frac{251\pm\sqrt{8137}}{1080}$ which has two allowed values
of $n$ which are approximately $n = 1/3,1/7$. In case $n = 1/3$, we find
$A^2$ is approximately $3.5$ while in case $n = 1/7$ then one finds that 
$A^2$ is approximately $1.1$.

\section{Appendix B: Periodic and Hyperbolic Solutions of the Hyperbolic
Plane Pendulum Equation in the First Anharmonic Approximation}

We start from the hyperbolic plane pendulum Eq. (\ref{20}). In the 
harmonic approximation, i.e. the case of small $\theta$, the hyperbolic 
pendulum Eq. (\ref{20}) is again given by Eq. (\ref{1.1}) whose solution is 
as given by Eq. (\ref{1.2}) so that in the harmonic approximation the 
hyperbolic plane pendulum is an example of an isochronous system. 

In the first anharmonic approximation, we have $\sinh(\theta) = \theta 
+\frac{\theta^3}{6}+...$, so that Eq. (\ref{20}) takes the form
\be\label{B1}
\theta_{tt} = -a\theta - \frac{a\theta^3}{6}-...\,.
\ee
We shall now obtain the solution of the hyperbolic pendulum Eq. (\ref{B1}) 
in the first anharmonic approximation. For clarity, we shall first obtain 
solutions of the equation
\be\label{B2}
\theta_{tt} = -a \theta - b \theta^3\,,~~a,b > 0\,,
\ee
where $a = {g}/{l}$ and then use $b = a/6$ to obtain a condition under which
it is a solution of the hyperbolic plane pendulum Eq. (\ref{B1}).

{\bf Solution I} \\
It is easy to check that 
\be\label{B3}
\theta(t) = A \sqrt{n} \cn(\beta t,n)\,,
\ee
is an exact solution of Eq. (\ref{B2}) provided $0 < n < 1/2$ and further
\be\label{B4}
b A^2 = 2 \beta^2\,,~~a = (1-2n)\beta^2\,.
\ee
Thus, it is a solution of the hyperbolic plane pendulum Eq. (\ref{B1}) in case
\be\label{B5}
A = \frac{2\sqrt{3}}{\sqrt{(1-2n)}}\,. 
\ee
This is a periodic solution with time period $T$ being 
\be\label{B6}
T(n) = 4K(n)\sqrt{\frac{12}{a A^2}}\,,
\ee
where $K(n)$ is the complete elliptic integral of the first kind \cite{as}.

{\bf Solution II} \\

Eq. (\ref{B2}) also admits a complex PT-invariant periodic pulse solution 
with $PT$ eigenvalue +$1$ \cite{ks1}
\be\label{B7}
\theta(t) = A \dn(\beta t, n) +i B \sqrt{n} \sn(\beta t, n)\,,
\ee
provided $1/2 < n \le 1$ and further
\be\label{B8}
B = \pm A\,,~~2bA^2 = \beta^2\,,~~a = \frac{(2n-1)\beta^2}{2}\,.
\ee
Thus, it is a solution of the hyperbolic plane pendulum Eq. (\ref{B1}) in case
\be\label{B9}
A = \frac{\sqrt{6}}{\sqrt{(2n-1)}}\,. 
\ee
This is a periodic solution with time period $T$ being 
\be\label{B10}
T(n) = 4K(n)\sqrt{\frac{6}{a A^2}}\,. 
\ee

{\bf Solution III} \\
Eq. (\ref{B2}) admits another complex PT-invariant periodic pulse solution  
with $PT$ eigenvalue +$1$  \cite{ks1}
\be\label{B11}
\theta(t) = \sqrt{n} [A\cn(\beta t, n) +i B \sn(\beta t, n)]\,,
\ee
provided 
\be\label{B12}
B = \pm A\,,~~2bA^2 = \beta^2\,,~~a = \frac{(2-n)\beta^2}{2}\,.
\ee
Thus, it is a solution of the hyperbolic plane pendulum Eq. (\ref{B1}) in case
\be\label{B13}
A = \frac{\sqrt{6}}{\sqrt{(2-n)}}\,. 
\ee
This is a periodic solution with time period $T$ being 
\be\label{B14}
T(n) = 4K(n)\sqrt{\frac{6}{a A^2}}\,. 
\ee

{\bf Solution IV}

In the limit $n = 1$, the solutions II and III  go over to the complex
hyperbolic pulse solution with PT-eigenvalue +$1$
\be\label{B15}
\theta(t) = A \sech(\beta t) +i B \tanh(\beta t)\,,
\ee
provided
\be\label{B16}
B = \pm A\,,~~2bA^2 = \beta^2\,,~~a = \frac{\beta^2}{2}\,.
\ee
Thus, it is a solution of the hyperbolic plane pendulum Eq. (\ref{B1}) in case
\be\label{B17}
A = \sqrt{6}\,.
\ee

{\bf Solution V}\\
Remarkably, we find that Eq. (\ref{B2}) also admits a 
real superposed pulse solution \cite{ks2}
\be\label{B18}
\theta(t) = A \dn(\beta t, n) + B \sqrt{m} \cn(\beta t, n)\,,
\ee
provided
\be\label{B19}
B = \pm A\,,~~ 2bA^2  = \beta^2\,,~~a = \frac{(1+n)\beta^2}{2}\,.
\ee
Thus, it is a solution of the hyperbolic plane pendulum Eq. (\ref{B1}) in case
\be\label{B20}
A = \frac{\sqrt{6}}{\sqrt{(1+n)}}\,. 
\ee
This is a periodic solution with time period $T$ being 
\be\label{B21}
T(n) = 4K(n)\sqrt{\frac{6}{a A^2}}\,. 
\ee

{\bf Solution VI} \\
Eq. (\ref{B2}) admits another periodic solution \cite{ks4}
\be\label{B22}
\theta(t) = \frac{A \sn(\beta t,n)}{1+B\cn^2(\beta t,n)}\,,~~B > 0\,,
\ee
provided $0 < n < 1/36$, and further
\bea\label{B23}
&&0 < n < 1\,,~~B = \frac{\sqrt{n}}{1-\sqrt{n}}\,, \nonumber \\
&&a = [(1+n)-6\sqrt{n}]\beta^2\,,~~b A^2 = 8\sqrt{n} \beta^2\,.
\eea
Thus, it is a solution of the hyperbolic plane pendulum Eq. (\ref{B1}) in case
\be\label{B24}
A = \frac{4\sqrt{3\sqrt{n}}}{\sqrt{(1+n-6\sqrt{n})}}\,. 
\ee
This is a periodic solution with time period $T$ being 
\be\label{B25}
T(n) = 4K(n)\sqrt{\frac{(1+n -6\sqrt{n})}{a}}\,. 
\ee

On using the identity \cite{as}
\be\label{B26}
\sn(y+\Delta,n)+\sn(y-\Delta,n) = \frac{2\sn(y,n) \cn(\Delta,n)}
{\dn(\Delta,n)[1+B\cn^2(y,n)]}\,,~~B = \frac{n\sn^2(\Delta,n)}
{\dn^2(\Delta,n)}\,,
\ee
one can re-express the solution (\ref{B22}) as a superposition of two 
periodic kinklike solutions
\be\label{B27}
\theta(t) = \frac{2\sqrt{n}}{a\sqrt{1+n-6\sqrt{n}}} 
 [\sn(\beta t +\Delta, n)+\sn(\beta t -\Delta, n)]\,.
\ee
Here $\Delta$ is defined by $\sn(\sqrt{n}\Delta,1/n) = \pm n^{1/4}$,
where use has been made of the identity (\ref{1.29}) \cite{as}.

{\bf Solution VII}

Eq. (\ref{B2}) also admits a complex PT-invariant periodic 
superposed solution with PT-eigenvalue $+1$
\be\label{B36}
\theta(t) = \frac{[A\dn(\beta t,n)+iB\sqrt{n}\sn(\beta t,n)]}
{D+\cn(\beta t,n)}\,,~~D > 1\,,
\ee
provided $1/2 < n \le 1$ and further
\be\label{B37}
2 b A^2 = (D^2-1)\beta^2\,,~~2 n b B^2 = (n D^2+1-n)\beta^2\,,~~
a = (2n-1)\frac{\beta^2}{2}\,.
\ee
Thus, it is a solution of the hyperbolic plane pendulum Eq. (\ref{B1}) in case
\be\label{B38}
A = \frac{\sqrt{6(D^2-1)}}{\sqrt{(2n-1)}}\,. 
\ee
This is a periodic solution with time period $T$ being 
\be\label{B39}
T(n) = 4K(n)\sqrt{\frac{6(D^2-1)}{a A^2}}\,. 
\ee

{\bf Solution VIII}

In the limit $n = 1$, the solution VII goes over to the hyperbolic complex 
PT-invariant superposed solution with PT-eigenvalue $+1$
\be\label{B40}
\theta(t) = \frac{[A\sech(\beta t)+iB \tanh(\beta t)]}{D+\sech(\beta t)}\,,
~~D > 1\,,
\ee
provided 
\be\label{B41}
2 b A^2 = (D^2-1)\beta^2\,,~~2 b B^2 = D^2 \beta^2\,,~~
a = \frac{\beta^2}{2}\,.
\ee
Thus, it is a solution of the hyperbolic plane pendulum Eq. (\ref{B1}) in case
\be\label{B42}
A = \sqrt{6(D^2-1)}\,. 
\ee

{\bf Solution IX}

Eq. (\ref{B2}) also admits a complex PT-invariant periodic superposed kink 
solution with PT-eigenvalue $-1$
\be\label{B43}
\theta(t) = \frac{\sqrt{n}[A\sn(\beta t,n)+iB\cn(\beta t,n)]}
{D+\dn(\beta t,n)}\,,~~0 < D^2 < 1-n\,,
\ee
provided
\be\label{B44}
2 b n A^2 = (1-D^2-n)\beta^2\,,~~2 b B^2 = (1-D^2)\beta^2\,,~~
a = (2-n)\frac{\beta^2}{2}\,.
\ee
Thus, it is a solution of the hyperbolic plane pendulum Eq. (\ref{B1}) in case
\be\label{B45}
A = \frac{\sqrt{6(1-D^2-n)}}{\sqrt{n(2-n)}}\,. 
\ee
This is a periodic solution with time period $T$ being 
\be\label{B46}
T(n) = 4K(n)\sqrt{\frac{6(1-D^2-n)}{n a A^2}}\,. 
\ee

{\bf Solution X}

Eq. (\ref{B2}) also admits a complex PT-invariant periodic 
superposed pulse solution with PT-eigenvalue $+1$
\be\label{B47}
\theta(t) = \frac{\sqrt{n}[A\cn(\beta t,n)+iB\sn(\beta t,n)]}
{D+\dn(\beta t,n)}\,,~~D^2 > 1\,,
\ee
provided
\be\label{B48}
2 b n B^2 = (D^2+n-1)\beta^2\,,~~2 b A^2 = (D^2-1)\beta^2\,,~~
a = (2-n)\frac{\beta^2}{2}\,.
\ee
Thus, it is a solution of the hyperbolic plane pendulum Eq. (\ref{B1}) in case
\be\label{B49}
A = \frac{\sqrt{6(1-D^2-n)}}{\sqrt{n(2-n)}}\,. 
\ee
This is a periodic solution with time period $T$ being 
\be\label{B50}
T(n) = 4K(n)\sqrt{\frac{6(1-D^2-n)}{n a A^2}}\,. 
\ee

{\bf Solution XI} \\
Eq. (\ref{B1}) admits another periodic superposed solution \cite{ks4}
\be\label{B51}
\theta(t) = \frac{A \sn(\beta t,n) \cn(\beta t,n)}{1+B\cn^2(\beta t,n)}\,,
\ee
provided
\bea\label{B52}
&&0 < n \le 24/25\,,~~B = \frac{1-\sqrt{1-n}}{\sqrt{1-n}}\,, 
\nonumber \\
&&a = (6\sqrt{1-n}-2+n)\beta^2\,,~~ b A^2 
= \frac{8(1-\sqrt{1-n})^2 \beta^2}{\sqrt{1-n}}\,.
\eea 
Thus, it is a solution of the hyperbolic plane pendulum Eq. (\ref{B1}) in case
\be\label{B53}
A = \frac{4(1-\sqrt{1-n})\sqrt{3}}{\sqrt{\sqrt{1-n}[6\sqrt{1-n}-(2-n)]}}\,. 
\ee
This is a periodic solution with time period $T$ being 
\be\label{B54}
T(n) = 2K(n)\sqrt{\frac{(48(1-\sqrt{1-n})^2}{\sqrt{1-n} a A^2}}\,. 
\ee

On using the identity \cite{as}
\bea\label{B55}
&&\dn(y-\Delta,n) -\dn(y+\Delta,n) = \frac{2n\sn(\Delta,n) \cn(\Delta,n) 
\sn(y,n) \cn(y,n)}{\dn^2(\Delta,n)[1+ B \cn^2(y,n)]}\,, \nonumber \\
&&B = \frac{n \sn^2(\Delta,n)}{\dn^2(\Delta,n)}\,,  
\eea
one can re-express solution (\ref{B51}) as a superposition of two 
periodic pulse-like solutions, i.e.
\be\label{B56}
\theta(t) = \frac{2}{a} \sqrt{\frac{3}{[6\sqrt{1-n}-(2-n)]}} 
\bigg(\dn[\beta t -\frac{K(n)}{2},n] - \dn[\beta t +\frac{K(n)}{2},n] \bigg)\,.
\ee

\section{Appendix C: Some Basic Results about Jacobi Elliptic Functions}

In this appendix we mostly mention those properties of the Jacobi elliptic
functions which have been used in this paper. For more details about these
functions, see \cite{as}. There are three basic Jacobi elliptic functions 
$\sn(x,m)$, $\cn(x,m)$ and $\dn(x,m)$, 
where $0 \le m \le 1$ is the modulus of the Jacobi elliptic functions. In the
special cases of $m = 0$ and $m = 1$, these Jacobi elliptic functions go over
to the trigonometric and the hyperbolic functions, respectively. In particular, 
\be\label{C1}
\sn(x,m=0) = \sin(x)\,,~~\cn(x,m=0) = \cos(x)\,,~~\dn(x,m=0) = 1\,,
\ee
\be\label{C2}
\sn(x,m=1) = \tanh(x)\,,~~\cn(x,m=1) = \dn(x,m=1) = \sech(x)\,.
\ee
The three Jacobi elliptic functions satisfy the identities
\be\label{C3}
\sn^2(x,m)+\cn^2(x,m) = 1 = \dn^2(x,m) + m \sn^2(x,m)\,.
\ee
While $\cn(x,m)$ and $\dn(x,m)$ are even functions of $x$, $\sn(x,m)$ is an 
odd function of $x$, i.e.
\bea\label{C4}
&&\sn(-x,m) = -\sn(x,m)\,,~~\cn(-x,m) = \cn(x,m)\,, \nonumber \\
&&\dn(-x,m) = \dn(x,m)\,.
\eea
Note that $\sn(x,m)$, $\cn(x,m)$, and $dn(x,m)$ are all doubly periodic functions. 
While the real period of $\sn(x,m)$ and $\cn(x,m)$ is $4K(m)$, that of $\dn(x,m)$ 
is $2K(m)$, i.e.
\bea\label{C5}
&&\sn[x+4K(m),m] = \sn(x,m)\,,~~\cn[x+4K(m),m] = \cn(x,m)\,, \nonumber \\
&&\dn[x+2K(m),m] = \dn(x,m)\,.
\eea
Some special values of these functions are
\bea\label{C6}
&&\sn(x=0,m) = \sn(x=2K(m),m) = 0\,, \nonumber \\
&&\sn(x=K(m),m) = -\sn(x=3K(m),m) = 1\,,
\eea
\bea\label{C7}
&&\cn(x=0,m) = -\cn(x=2K(m),m) = 1\,, \nonumber \\
&&\cn(x=K(m),m) = \cn(x=3K(m),m) = 0\,,
\eea
\be\label{C8}
\dn(x=0,m) = 1\,,~~\dn(x=K(m),m) = \sqrt{1-m}\,. 
\ee
Thus, $\dn(x,m)$ is nonnegative and oscillates between $\sqrt{1-m}$ 
and $1$. On the other hand, $\sn(x,m)$ and $\cn(x,m)$ go from $-1$ to $1$.
On differentiating these three functions with respect to 
their argument $x$ we obtain
\bea\label{C9}
&&\frac{d\sn(x,m)}{dx} = \cn(x,m) \dn(x,m)\,,~~\frac{d\cn(x,m)}{dx} = -\sn(x,m)
\dn(x,m)\,, \nonumber \\
&&\frac{d\dn(x,m)}{dx} = -m \sn(x,m) \cn(x,m)\,.
\eea
The addition theorems for the three Jacobi elliptic functions are
\be\label{C10}
\sn(x+y,m) = \frac{\sn(x,m)\cn(y,m)\dn(y,m)+\sn(y,m)\cn(x,m)\dn(x,m)}{1-
m\sn^2(x,m)\sn^2(y,m)}\,,
\ee
\be\label{C11}
\cn(x+y,m) = \frac{\cn(x,m)\cn(y,m)-\sn(x,m)\sn(y,m)\dn(x,m)\dn(y,m)}{1-
m\sn^2(x,m)\sn^2(y,m)}\,,
\ee
\be\label{C12}
\dn(x+y,m) = \frac{\dn(x,m)\dn(y,m)-m\sn(x,m)\sn(y,m)\cn(x,m)\cn(y,m)}{1-
m\sn^2(x,m)\sn^2(y,m)}\,. 
\ee

\end{document}